\documentstyle[12pt]{article}

\begin{document}

\begin{center}
{\bf WAVE EQUATION SOLUTIONS AND PAIR PRODUCTION FOR ARBITRARY 
SPIN PARTICLES}\\
\vspace{5mm} {\bf S.I.Kruglov}
\footnote{E-mail: krouglov@sprint.ca} \\
\vspace{5mm}
{\it International Educational Centre, 2727 Steeles Ave.West, Suite 202, \\
Toronto, Ontario, Canada M3J 3G9}\\
\vspace{5mm}
\end{center}

\begin{abstract}
We investigate the theory of particles with arbitrary spin and
magnetic moment in the Lorentz representation $\left(0,s\right)
\oplus \left( s,0\right)$ in an external constant and uniform
electromagnetic field. We obtain the density matrix of free
particles in pure spin states. The differential probability of
pair producing particles with arbitrary spin by an external
constant and uniform electromagnetic field is found using the
exact solutions. We calculate the imaginary and real parts of the
Lagrangian in an electromagnetic field that takes into account the
vacuum polarization.
\end{abstract}

\section{ Introduction}

Interest in the theory of relativistic particles with arbitrary
spins is increasing. One of the reasons is that SUSY models
require superpartners, i.e., additional fields of particles with
higher spins. In particular, it is important to take into account
particle with spin 3/2-gravitino which appears in cosmological
models based on supergravity and in the theory of inflation of the
universe [1].

Another factor is that some string models have similar features to
models of relativistic spinning particles [2]. It is also
interesting to have particles with arbitrary fractional spins [3]
(see also refs. [4,5]). Such spinning particles in $(2+1)$
dimensions, called anyons, were discovered and have anomalous
statistics.

There are many different relativistic wave equations which
describe particles with arbitrary spins [6-11]. The fields of free
particles of a mass $m$ and spin $s$ in these formalisms realize
definite representations of the Poincar$\acute{e}$ group. Some of
these schemes are equivalent to each other. If the interaction
with external electromagnetic fields is introduced, approaches
based on different representations of the Lorentz group become
inequivalent. Most theories of particles in external
electromagnetic fields have difficulties such as non-causal
propagation [12], indefinite metrics in the second-quantized
theory [13-15], etc.

We proceed from the second-order relativistic wave equation for
particles with arbitrary spin $s$ and magnetic moment $\mu$, on
the basis of the Lorentz representation $(0,s)\oplus (s,0)$, where
$\oplus$ means the direct sum. In this scheme the Hermitianizing
matrix $\eta $ exists for arbitrary spin, as there are pairs of
mutually conjugate representations $ \left(0,s\right)$ and
$\left(s,0\right)$ [11]. Therefore we have here a Lagrangian
formulation and the scalar product of wave functions, which is
important for any quantum mechanical calculations. Such an
approach avoids difficulties of other schemes and the
corresponding wave function has the minimal number of components.
This is a generalization of the Feynman-Gell-Mann equation [16]
for particles with spin $1/2$ to the case of higher spin particles
which possess arbitrary magnetic moment. In the particular case of
spin $1/2$ we recover to the Dirac theory. If the normal magnetic
moment is considered, it leads to the approach of refs. [11].
Particles in this scheme propagate causally in external
electromagnetic fields and this is a parity-symmetric theory with
a Lagrangian formulation. Some equations on the basis of
$(0,s)\oplus (s,0)$ representations of the Lorentz group were
studied in refs. [17]. In refs. [11], the $2(6s+1)$ -component,
first-order matrix parity-invariant formulation of the equation
for particles with arbitrary spin was considered. Then the author
obtained the second-order equation for particles with ``normal''
magnetic moment. Starting with the second-order equation for
particles which possess arbitrary magnetic moment, we go to the
first-order wave equation with another representation of the
Lorentz group; the algebraic properties of the matrices are the
same as in the approach of refs. [11]. We find solutions of
equations for free particles in the form of density matrices
(projection matrix-dyads) for pure spin states which are used for
different electromagnetic evaluation of the Feynman diagrams. Such
projection matrix-dyads allow us to make covariant calculations
without using matrices of the wave equation in a definite
representation.

The main purpose of this paper is to investigate solutions of wave
equations, pair production of arbitrary-spin particles by
constant, uniform electromagnetic fields, and vacuum polarization
of higher spin particles. Considering one-particle theory, and
obtaining the differential probability for pair production of
particles with arbitrary spins, we avoid the Klein paradox [18,
19]. As a particular case of spin-$1/2$ and gyromagnetic ratio $2$
particles, we arrive at the well-known result found by Schwinger
[20], who predicted $e^{+}e^{-}$ pair production in a strong
external electromagnetic field. This has now been realized by the
development of power laser techniques. It should be noted that the
pair production of particles by a gravitational field is also
important for understanding the evolution of the universe near a
singularity [21].

The probability of pair production of particles in external
electromagnetic fields can be found on the basis of exact
solutions of the wave equations [22] or the imaginary part of the
Lagrangian [20]. We consider here both approaches. Nonlinear
corrections to the Maxwell Lagrangian of the constant uniform
electromagnetic fields are determined from the polarization of the
vacuum of arbitrary-spin particles. The problem of pair production
of particles with higher spins using the quasiclassical scheme
(method of ``imaginary time'') was considered in ref. [23] and is
in accord with our approach via relativistic wave equations. It
should be noted that the quasiclassical approximation has a
restriction for the fields $E,$ $H\ll m^2/e$ when the process is
exponentially suppressed. It means that the approach of ref. [23]
is valid when the electromagnetic fields are not too strong, i.e.,
less than the critical value $m^2/e$. But it is known that pairs
of particles are created rapidly at this critical value of the
fields. In our consideration, there are no such restrictions. The
problem of the pair production of vector particles with
gyromagnetic ratio $2$ was investigated in ref. [24]. In refs.
[25, 26], the imaginary part of the effective Lagrangian which
defines the probability of $e^{+}e^{-}$-production was found by
taking into account the anomalous magnetic moment.

We use system of units $\hbar =c=1$, $\alpha =e^2/4\pi =1/137$,
$e>0$. In Section 2, proceeding from the second-order equation for
arbitrary-spin particles with anomalous magnetic moment, we go to
the first-order formulation of the theory. All independent
solutions of the equation for free particles are found in the form
of matrix-dyads (density matrices). In Section 3, we consider the
important case of spin-1 particles, which is different from the
Proca or Petiau-Duffin-Kemmer theories. Here, the $2(6s+1)$
-component first-order equation is constructed. Section 4
investigates exact solutions of arbitrary-spin particle equations
in constant, uniform electromagnetic fields. On the basis of exact
solutions, we find the differential probability of pair production
of particles with arbitrary spin and anomalous magnetic moment.
The imaginary part of the effective Lagrangian for electromagnetic
fields is calculated. In Section 5, using the Schwinger method, we
find the nonlinear corrections to the Lagrangian of a constant,
uniform electromagnetic field caused by the vacuum polarization of
particles with arbitrary spin and magnetic moment. Section 6
discusses of results.

\section{ Wave Equation and Density Matrix}

We proceed here from the theory based on the $(s,0)\oplus (0,s)$
Lorentz representation for massive particles. The wave function of
the $(s,0)\oplus (0,s)$ representation has $2(2s+1)$ components.
For spin $1/2$, we arrive at well-known Dirac bispinors. For spin
$1$, however, there is doubling of the component compared with the
Proca theory [27] because the vector particles have three spin
states with the projections $s_z=\pm 1,0$.

We postulate the following two (for $\varepsilon =\pm 1$) wave
equations for arbitrary-spin particles in external electromagnetic
fields:

\begin{equation}
\left( D_\mu ^2-m^2-\frac{eq}{2s}F_{\mu \nu }\Sigma _{\mu \nu
}^{(\varepsilon )}\right) \Psi _\varepsilon (x)=0, \label{1}
\end{equation}
where $s$ is the spin of particles, $D_\mu =\partial _\mu -ieA_\mu
$; $ F_{\mu \nu }=\partial _\mu A_\nu -\partial _{_\nu }A_\mu $ is
the strength tensor of external electromagnetic fields,
$\varepsilon =\pm 1;$ and $\Sigma _{\mu \nu }^{\left( -\right) }$,
$\Sigma _{\mu \nu }^{\left( +\right) }$ are the generators of the
Lorenz group which correspond to the $\left( s,0\right) $ and
$\left( 0,s\right) $ representations. Two equations (1) (for
$\varepsilon =\pm 1$) describe particles which possess the
magnetic moment $\mu =eq/(2m)$ and gyromagnetic ratio $g=q/s$. At
$q=1$, we have the ``normal'' magnetic moment $\mu =e/(2m)$ and
$g=1/s$. The generators $\Sigma _{\mu \nu }^{(\varepsilon )}$ are
connected with the spin matrices $S_k$ by the relationships
$\Sigma _{ij}^{(\varepsilon )}=\epsilon _{ijk}S_k$ and $ \Sigma
_{4k}^{(\varepsilon )}=-i\varepsilon S_k$, where the parameter $
\varepsilon $ corresponds to the Lorentz group representations
$\left( s,0\right) $ for $\varepsilon =1$ and $\left( 0,s\right) $
for $\varepsilon =-1$. As usual, relations

\begin{equation}
\left[ S_i,S_j\right] =i\epsilon _{ijk}S_k,\hspace{1.0in}\left( S_1\right)
^2+\left( S_2\right) ^2+\left( S_3\right) ^2=s(s+1) \label{2}
\end{equation}
are valid, where $i,j,k=1,2,3; \epsilon _{ijk}$ is the Levi-Civita
symbol ($ \epsilon _{123}=1$).

At $q=1$ equations (1) where considered in refs. [11]. The theory
of arbitrary spin particles based on Eq. (1) is causal in the
presence of external electromagnetic fields. It is seen from the
method of ref. [12] that the equations remain hyperbolic and the
characteristic surfaces are lightlike. Equation (1) is invariant
under the parity operation. Indeed, at the parity inversion $
\varepsilon \rightarrow -\varepsilon $, the representation $(s,0)$
transforms into $(0,s)$.

Now we will consider the problem of formulating a first-order
relativistic wave equation from the second-order equation (1).
This is convenient for some quantum electrodynamic calculations
with polarized particles of arbitrary spins.

Let us introduce the matrix $\varepsilon ^{A,B}$ with dimension $
n\times n$; its elements consist of zeroes and only one element is
unity, where row $A$ and column $B$ cross. So the matrix elements
and multiplication of this matrices are

\begin{equation}
\left( \varepsilon ^{A,B}\right) _{CD}=\delta _{AC}\delta _{BD},
\hspace{1.0in}\varepsilon ^{A,B}\varepsilon ^{C,D}=\delta _{BC}\varepsilon
^{A,D}, \label{3}
\end{equation}
where indexes $A,B,C,D=1,2...n$.

Six generators $\Sigma _{\mu \nu }^{(+)}$ (or $\Sigma _{\mu \nu
}^{(-)}$) occurring in Eq. (1) have the dimension $2s+1$.
Therefore the wave function $\Psi _{+}(x)$ (or $\Psi _{-}(x)$) of
Eq. (1) possesses $2s+1$ components. Now we can introduce the
$5(2s+1)$-component wave function

\begin{equation}
\Psi _1(x)=\left(
\begin{array}{c}
\Psi _{+}(x) \\
-\frac 1mD_\mu \Psi _{+}(x)
\end{array}
\right) \label{4}
\end{equation}
so that $\Psi _1(x)=\left\{ \Psi _A(x)\right\},$ $A=0,\mu ;$ $\Psi
_0=\Psi _{+}(x),$ $\Psi _\mu =-(1/m)D_\mu \Psi _{+}(x)$ and $\Psi
_{+}(x)$ realizes the Lorentz representation $(s,0)$. It is not
difficult to check that Eq. (1) for $\varepsilon =+1$ can be
represented as the first-order equation

\begin{equation}
\left( \beta _\mu ^{(+)}D_\mu +m\right) \Psi _1(x)=0, \label{5}
\end{equation}
where $5(2s+1)\times 5(2s+1)$- matrices

\begin{equation}
\beta _\mu ^{(+)}=\left( \varepsilon ^{0,\mu }+\varepsilon ^{\mu ,0}\right)
\otimes I_{2s+1}-i\frac qs\varepsilon ^{0,\nu }\otimes \Sigma _{\mu \nu
}^{(+)}, \label{6}
\end{equation}
are introduced, and $\otimes $ is the direct product of matrices,
$I_{2s+1}$ is a unit matrix of the dimension $2s+1$, and in (6),
we imply the summation on index $\nu $. It should be noted that in
refs. [11], the $\left( 6s+1\right) $ -component wave function was
introduced for the case of ``normal'' magnetic moment of
particles. Here the higher dimension $5\left( 2s+1\right) $ of the
wave function is considered as compared with the $\left(
6s+1\right) $ -component function investigated in refs. [11]. The
difference is that we do not introduce the vector potentials for
arbitrary spin fields and deal only with the strength field tensor
$\Psi _{+}(x)$ and its derivatives $D_\mu \Psi _{+}(x)$. The
particular case of a spin-1 field will be considered in Section 3
on the basis of $\left( 6s+1\right) $-representation of the wave
function by introducing vector potentials. Using properties (3),
it is easy to check that the five-dimension matrices $\beta _\mu
^{PDK}=\varepsilon ^{0,\mu }+\varepsilon ^{\mu ,0}$ obey the
Petiau-Duffin-Kemmer algebra [28-30],

\begin{equation}
\beta _\mu ^{PDK}\beta _\nu ^{PDK}\beta _\alpha ^{PDK}+\beta _\alpha
^{PDK}\beta _\nu ^{PDK}\beta _\mu ^{PDK}=\delta _{\mu \nu }\beta _\alpha
^{PDK}+\delta _{\alpha \nu }\beta _\mu ^{PDK}. \label{7}
\end{equation}

The wave function $\Psi _1(x)$ transforms on the $\left[ \left(
0,0\right) \oplus \left( 1/2,1/2\right) \right] \otimes (s,0)$
representation of the Lorentz group [31,32]. For the case
$\varepsilon =-1$, we have the analogous equation

\begin{equation}
\left( \beta _\mu ^{(-)}D_\mu +m\right) \Psi _2(x)=0, \label{8}
\end{equation}
where
\[
\beta _\mu ^{(-)}=\left( \varepsilon ^{\overline{0},\overline{\mu }
}+\varepsilon ^{\overline{\mu },\overline{0}}\right) \otimes I_{2s+1}-i\frac
qs\varepsilon ^{\overline{0},\overline{\nu }}\otimes \Sigma _{\mu \nu
}^{(-)},
\]

\begin{equation}
\Psi _2(x)=\left(
\begin{array}{c}
\Psi _{-}(x) \\
-\frac 1mD_\mu \Psi _{-}(x)
\end{array}
\right) \label{9}
\end{equation}
and $\Psi _2(x)=\left\{ \Psi _B(x)\right\} $,
$B=\overline{0},\overline{\mu } ;$ $\Psi _{\overline{0}}(x)=\Psi
_{-}(x)$, $\Psi _{\overline{\mu }}=-\frac 1mD_\mu \Psi _{-}(x)$,
where $\Psi _{-}(x)$ transforms as $(0,s)$ representation of the
Lorentz group, and $\Psi _2(x)$ realizes the $\left[ \left(
0,0\right) \oplus \left( 1/2,1/2\right) \right] \otimes (0,s)$
representation. The two equations (5) and (8) can be combined into
one first-order equation

\begin{equation}
\left( \beta _\mu D_\mu +m\right) \Psi (x)=0 \label{10}
\end{equation}
with the matrices and wave function

\begin{equation}
\beta _\mu =\beta _\mu ^{(+)}\oplus \beta _\mu ^{(-)},\hspace{1.0in}\Psi
(x)=\left(
\begin{array}{c}
\Psi _1(x) \\
\Psi _2(x)
\end{array}
\right). \label{11}
\end{equation}

Using the properties of elements of the entire algebra (3), it is
not difficult to verify that the matrices $\beta _\mu $ (the same
for $\beta _\mu ^{(+)}$ and $\beta _\mu ^{(-)}$) obey the algebra
\[
\beta _\mu \beta _\nu \beta _\sigma +\beta _\nu \beta _\sigma \beta _\mu
+\beta _\sigma \beta _\mu \beta _\nu +\beta _\nu \beta _\mu \beta _\sigma
+\beta _\mu \beta _\sigma \beta _\nu +\beta _\sigma \beta _\nu \beta _\mu =
\]

\begin{equation}
=2\left( \delta _{\nu \sigma }\beta _\mu +\delta _{\mu \sigma }\beta _\nu
+\delta _{\mu \nu }\beta _\sigma \right). \label{12}
\end{equation}

In other refs. [11], the $2(6s+1)$-dimensional representation of
the $SL(2,C)$ group was considered for particles of arbitrary spin
with algebra (12). Here we study another representation of the
$SL(2,C)$ group, and the matrices $\beta _\mu$ of (11) are
$10\left( 2s+1\right) \times 10\left( 2s+1\right) $ dimensional.
Different representations of this algebra were studied in refs.
[33] and [34].

Let us consider the problem of finding the solutions to Eq. (10)
for definite momentum and spin projection. It is convenient to
find these solutions in the form of projection matrix-dyads
(density matrices). All electrodynamic calculations of Feynman
diagrams with arbitrary-spin particles can be done using these
matrices. As particles in initial and final states are free
particles, we can put the parameter $q=0$ in (1), (10). This
corresponds to the case when external electromagnetic fields are
absent. Then the matrices $\beta _\mu $ transform to $\beta _\mu
^0:$

\begin{equation}
\beta _\mu ^0=\left[ \left( \varepsilon ^{0,\mu }+\varepsilon ^{\mu
,0}\right) \otimes I_{2s+1}\right] \oplus \left[ \left( \varepsilon ^{
\overline{0},\overline{\mu }}+\varepsilon ^{\overline{\mu },\overline{0}
}\otimes I_{2s+1}\right) \right], \label{13}
\end{equation}
which obey the Petiau-Duffin-Kemmer algebra (7). The projection
operators extracting states with definite 4-momentum $p_\mu $ for
particle and antiparticle are given by

\begin{equation}
M_{\pm }=\frac{i\widehat{p}\left( i\widehat{p}\pm m\right) }{2m^2},
\label{14}
\end{equation}
where $\widehat{p}=p_\mu \beta _\mu ^0$ (we use the metric such
that $p^2= {\bf p}^2+p_4^2={\bf p}^2-p_0^2=-m^2$). Plus and minus
signs in (14) correspond to the particle and antiparticle,
respectively. Matrices $\Lambda _{\pm }$ have the usual projection
operator property [35]

\begin{equation}
M _{\pm }^2=M _{\pm }. \label{15}
\end{equation}

Equation (15) is verified by the relation
$\widehat{p}^3=p^2\widehat{p}$, which follows from the
Petiau-Duffin-Kemmer algebra (7). To find the spin projection
operators, we need the generators of the Lorentz group in the
representation of the wave function $\Psi (x)$ which occurs in Eq.
(10). From the structure of the functions $\Psi _1(x)$ and $\Psi
_2(x)$ in (4) and (9), we define the generators of the Lorentz
group in our $10(2s+1)$-dimensional representation
\[
J_{\mu \nu }=J_{\mu \nu }^{(+)}\oplus J_{\mu \nu }^{(-)},
\]

\begin{equation}
J_{\mu \nu }^{(+)}=\left( \varepsilon ^{\mu ,\nu }-\varepsilon ^{\nu ,\mu
}\right) \otimes I_{2s+1}+iI_5\otimes \Sigma _{\mu \nu }^{(+)}, \label{16}
\end{equation}

\[
J_{\mu \nu }^{(-)}=\left( \varepsilon ^{\overline{\mu },\overline{\nu }
}-\varepsilon ^{\overline{\nu },\overline{\mu }}\right) \otimes
I_{2s+1}+iI_5\otimes \Sigma _{\mu \nu }^{(-)},
\]
where $I_5$ is the $5$-dimension unit matrix. Using properties
(3), we get the commutation relations for $J_{\mu \nu }$, and

\begin{equation}
\left[ J_{\mu \nu },J_{\alpha \beta }\right] =\delta _{\nu \alpha }J_{\mu
\beta }+\delta _{\mu \beta }J_{\nu \alpha }-\delta _{\mu \alpha }J_{\nu
\beta }-\delta _{\nu \beta }J_{\mu \alpha }, \label{17}
\end{equation}

\begin{equation}
\left[ \beta _\mu ,J_{\alpha \beta }\right] =\delta _{\mu \alpha }\beta
_\beta -\delta _{\mu \beta }\beta _\alpha. \label{18}
\end{equation}

Relationship (17) is a well-known commutation relation for
generators of the Lorentz group [31,32]. Equation (10) is a form
invariant under the Lorentz transformations because relation (18)
is valid. To guarantee the existence of a relativistically
invariant bilinear form

\begin{equation}
\overline{\Psi }(x)\Psi (x)=\Psi ^{+}(x)\eta \Psi (x), \label{19}
\end{equation}
where $\Psi ^{+}(x)$ is the Hermitian-conjugate wave function, we
should construct a Hermitianizing matrix $\eta $ with the
properties [33,35]:

\begin{equation}
\eta \beta _i=-\beta _i\eta ,\hspace{1.0in}\eta \beta _4=\beta _4\eta
\hspace{1.0in}(i=1,2,3). \label{20}
\end{equation}

Such a matrix exists because here there are two $\left( 0,s\right) $ and $
\left( s,0\right) $ mutually conjugate representations; $\eta$ is given by

\begin{equation}
\eta =\left( \varepsilon ^{a,\overline{a}}+\varepsilon ^{\overline{a}
,a}-\varepsilon ^{4,\overline{4}}-\varepsilon ^{\overline{4},4}-\varepsilon
^{0,\overline{0}}-\varepsilon ^{\overline{0},0}\right) \otimes I_{2s+1},\label{21}
\end{equation}
where the summation on the index $a=1,2,3$ is implied. Now we
introduce the operator of the spin projection on the direction of
the momentum ${\bf p}$:

\begin{equation}
\widehat{S}_{{\bf p}}=-\frac i{2\mid {\bf p\mid }}\epsilon
_{abc}p_aJ_{bc}=\left( \kappa _p+\sigma _p\right) \oplus \left( \overline{
\kappa }_p+\overline{\sigma }_p\right), \label{22}
\end{equation}
where
\[
\kappa _p=-\frac i{\mid {\bf p\mid }}\epsilon _{abc}p_a\varepsilon
^{b,c}\otimes I_{2s+1},\hspace{1.0in}\overline{\kappa }_p=-\frac i{\mid {\bf
p\mid }}\epsilon _{abc}p_a\varepsilon ^{\overline{b},\overline{c}}\otimes
I_{2s+1},
\]

\begin{equation}
\sigma _p=\overline{\sigma }_p=I_5\otimes \frac{{\bf pS}}{\mid
{\bf \ p\mid } }, \label{23}
\end{equation}
and $\mid {\bf p\mid =}\sqrt{p_1^2+p_2^2+p_3^2}$. It is easy to
check that the commutation relation holds:
\[
\left[ \widehat{S}_{{\bf p}},\widehat{p}\right] =0.
\]

The matrices $\kappa _p$, $\overline{\kappa }_p$ obey the simple equations

\begin{equation}
\kappa _p^3=\kappa _p,\hspace{1.0in}\overline{\kappa
}_p^3=\overline{\kappa } _p. \label{24}
\end{equation}

Taking into account Eqs. (2), we derive relations for the matrices
$ \sigma _p$ (23):
\[
\left( \sigma _p^2-\frac 14\right) \cdot \cdot \cdot \left( \sigma
_p^2-s^2\right) =0\hspace{1.0in} \mbox{for half-integer spins},
\]

\begin{equation}
\sigma _p\left( \sigma _p^2-1\right) \cdot \cdot \cdot \left(
\sigma _p^2-s^2\right) =0\hspace{1.0in} \mbox{for integer spins}.
\label{25}
\end{equation}

Relations (25) allow us to construct projection operators which extract the
pure spin states. Using the relationship
\[
\widehat{S}_{{\bf p}}\widehat{p}=\left( \sigma _p\oplus
\overline{\sigma } _p\right) \widehat{p},
\]
we can consider projection matrices on the basis of equations
(25). The common technique of the construction of such operators
is described in refs. [35]. Let us consider the equation for the
auxiliary spin operators $\sigma _p$, $ \overline{\sigma }_p$ for
the spin projection $s_k$:

\begin{equation}
\sigma _p\Psi _k=s_k\Psi _k. \label{26}
\end{equation}

The solution to Eq. (26) can be found using relationships (25),
which can be rewritten as

\begin{equation}
\left( \sigma _p-s_k\right) P_k(\sigma _p)=0, \label{27}
\end{equation}
where the polynomials $P_k(\sigma _p)$ are given by

\[
P_k(\sigma _p)=\left( \sigma _p^2-\frac 14\right) \cdot \cdot
\cdot \left( \sigma _p+s_k\right) \cdot \cdot \cdot \left( \sigma
_p^2-s^2\right) \hspace{0.3in} \mbox{for half-integer spins},
\]
\begin{equation}
P_k(\sigma _p)=\sigma _p\left( \sigma _p^2-1\right) \cdot \cdot
\cdot \left( \sigma _p+s_k\right) \cdot \cdot \cdot \left( \sigma
_p^2-s^2\right) \hspace{0.3in} \mbox{for integer spins}.
\label{28}
\end{equation}

Every column of the polynomial $P_k(\sigma _p)$ can be considered
as an eigenvector $\Psi_k$ of Eq. (26) with the eigenvalue $s_k$.
As $s_k$ is one multiple root of Eqs. (25), all columns of the
matrix $P_k(\sigma_p)$ are linearly independent solutions of Eq.
(26) [35]. It can be verified using definitions (28) that the
matrix

\begin{equation}
Q_k=\frac{P_k(\sigma _p)}{P_k(s_k)} \label{29}
\end{equation}
is the projection operator with the relation

\begin{equation}
Q_k^2=Q_k. \label{30}
\end{equation}

Equation (30) tells that the matrix $Q_k$ can be transformed into diagonal
form, with the diagonal containing only ones and zeroes. So the $Q_k$
acting on the wave function $\Psi $ will retain components which correspond
to the spin projection $s_k.$

We have mentioned that this theory of arbitrary-spin particles
involves the doubling of the spin states of particles because
there are two representations $(s,0)$ and $(0,s)$ of the Lorentz
group. To separate these representations, which are connected by
the parity transformations, we use the parity operator

\begin{equation}
K=\left( \varepsilon ^{\mu ,\overline{\mu }}+\varepsilon
^{\overline{\mu } ,\mu }+\varepsilon ^{0,\overline{0}}+\varepsilon
^{\overline{0},0}\right) \otimes I_{2s+1} \label{31}
\end{equation}
with the summation on index $\mu =1,2,3,4.$ The
$10(2s+1)$-dimensional matrix $ K$ has the property
$K^2=I_{10(2s+1)}.$ The projection operator extracting states with
the definite parity is given by

\begin{equation}
\Lambda _\varepsilon =\frac 12\left( K+\varepsilon \right), \label{32}
\end{equation}
where $\varepsilon =\pm 1$. This matrix possesses the required
relationship

\begin{equation}
\Lambda _\varepsilon ^2=\Lambda _\varepsilon. \label{33}
\end{equation}

It should be noted that the matrix $K$ (31) plays the role
analogous to the $ \gamma _5$ matrix in the Dirac theory of
particles with the spin $1/2$. It is easily verified that the
operators $\widehat{p},$ $\widehat{S}_{{\bf p}}$, $K$ commute with
each other, and, as a consequence, they have common eigenvectors.
The projection operator extracting pure states with definite
4-momentum projection, spin, and parity is given by

\begin{equation}
\Pi _{\pm m,k,\varepsilon }=M _{\pm }\Lambda_\varepsilon \left(
Q_k\oplus Q_k\right) \label{34}
\end{equation}
with matrices (14), (29) and (32). The $\Pi _{\pm m,k,\varepsilon
}$ is the density matrix for pure states. It is easy to consider
impure (mixed) states by summation of (34) over definite quantum
numbers $s_k$, $\varepsilon$. The projection operator for pure
states can be represented as matrix-dyad [36]

\begin{equation}
\Pi _{\pm m,k,\varepsilon }=\Psi _{\pm m,k}^\varepsilon \cdot
\overline{\Psi }_{\pm m,k}^\varepsilon, \label{35}
\end{equation}
where $\overline{\Psi }_{\pm m,k}^\varepsilon =\left( \Psi _{\pm
m,k}^\varepsilon \right) ^{+}\eta $, and $\Psi _{\pm
m,k}^\varepsilon $ is the solution to equations
\[
\left( i\beta _\mu p_\mu \pm m\right) \Psi _{\pm m,k}^\varepsilon =0,
\hspace{1.0in}\widehat{S}_{{\bf p}}\Psi _{\pm m,k}^\varepsilon =s_k\Psi
_{\pm m,k}^\varepsilon,
\]

\begin{equation}
K\Psi _{\pm m,k}^\varepsilon =\varepsilon \Psi _{\pm
m,k}^\varepsilon. \label{36}
\end{equation}

Expression (35) is convenient for calculations for different quantum
electrodynamics processes with polarized particles of arbitrary spins.

\section{ Wave Equation for Spin-1 Particles}

As a particular case, we consider the wave equation for particles
with spin 1. To compare our approach with refs. [11], the $2\left(
6s+1\right) $ representation for the wave function will be
studied. Equation (1) in the case of spin-1 particles becomes
\begin{equation}
\left( D_\lambda ^2-m^2\right) \Psi _{\mu \nu }(x)-ieq\delta
_{[\mu \nu ][\alpha \beta ]}F_{\alpha \sigma }\Psi _{\sigma \beta
}(x)=0, \label{37}
\end{equation}
where wave function $\Psi _{\mu \nu }(x)$ is the antisymmetric
tensor of the second rank $\Psi _{\mu \nu }(x)=-\Psi _{\nu \mu
}(x)$, which has six independent components, and $\delta _{[\mu
\nu ][\alpha \beta ]}=\delta _{\mu \alpha }\delta _{\nu \beta
}-\delta _{\mu \beta }\delta _{\nu \alpha }$. For comparison with
refs. [11], the ``normal'' magnetic moment $\mu =e/2m$ will be
considered here. Then the parameter $q=1$, and gyromagnetic ratio
$g=1$. Let us introduce two 4-vector potentials $\psi _\mu (x)$
and $\widetilde{\psi } _\mu (x)$ in accordance with the
relationship [36]
\begin{equation}
\Psi _{\mu \nu }(x)=D_\mu \psi _\nu (x)-D_\nu \psi _\mu
(x)-\epsilon _{\mu \nu \alpha \beta }D_\alpha \widetilde{\psi
}_\beta (x), \label{38}
\end{equation}
where $\epsilon _{\mu \nu \alpha \beta }$ is an antisymmetric
tensor Levi-Civita, $\epsilon _{1234}=-i$. This is a more general
representation of the antisymmetric tensor $\Psi _{\mu \nu }(x)$
via 4-vector $\psi _\mu (x)$ and 4-pseudovector $\widetilde{\psi
}_\mu (x)$. It is not difficult to verify that Eq. (37) (at $q=1$)
is a consequence of equations
\begin{equation}
D_\nu \Psi _{\mu \nu }(x)+m^2\psi _\mu (x)=0, \label{39}
\end{equation}
\begin{equation}
D_\nu \widetilde{\Psi }_{\mu \nu }(x)+m^2\widetilde{\psi }_\mu (x)
=0, \label{40}
\end{equation}
where $\widetilde{\Psi }_{\mu \nu }(x)=\frac 12\epsilon _{\mu \nu
\alpha \beta }\Psi _{\mu \nu }(x)$ is the dual tensor. So, the
system of the first-order equations (38)-(40) is equivalent to the
second-order equation (37). A doubling of spin states of particles
is obvious here. It follows from Eqs. (39) and (40) for free
particles (when external electromagnetic fields are absent and
$D_\mu =\partial _\mu -ieA_\mu \rightarrow \partial _\mu $) that
the Lorentz equations $\partial _\mu \psi _\mu (x)=0$, $
\partial _\mu \widetilde{\psi }_\mu (x)=0$ hold. Therefore the vector
potential $\psi_\mu (x)$ describes spin-1 states (the state with
spin 0 is not present due to the Lorentz equation) and
pseudovector $\widetilde{\psi }_\mu (x)$ also describes spin-1
states. As a result, there are six spin states and the system of
equations (38)-(40) is not equivalent to the Proca theory [27]. If
we set $\widetilde{\psi }_\mu (x)=0$ in (38)-(40), we will arrive
at the Proca equations.

Equations (38)-(40) are equivalent to the following system:
\[
D_\nu M_{\mu \nu }+m^2M_\mu =0,
\]
\begin{equation}
M_{\mu \nu }=D_\mu M_\nu -D_\nu M_\mu -\epsilon _{\mu \nu \alpha
\beta }D_\alpha M_\beta, \label{41}
\end{equation}
with the self-dual tensor $M_{\mu \nu }=i\widetilde{M}_{\mu \nu
}$, and
\[
D_\nu N_{\mu \nu }+m^2N_\mu =0,
\]
\begin{equation}
N_{\mu \nu }=D_\mu N_\nu -D_\nu N_\mu +\epsilon _{\mu \nu \alpha
\beta }D_\alpha N_\beta, \label{42}
\end{equation}
with the anti-self-dual tensor $N_{\mu \nu }=-i\widetilde{N}_{\mu
\nu }$, where
\begin{eqnarray*}
M_\mu &=&\frac 12\left( \psi _\mu (x)-i\widetilde{\psi }_\mu (x)\right),
\hspace{0.3in}M_{\mu \nu }=\frac 12\left( \Psi _{\mu \nu }(x)+i \widetilde{
\Psi }_{\mu \nu }(x)\right) , \\
N_\mu &=&\frac 12\left( \psi _\mu (x)+i\widetilde{\psi }_\mu (x)\right),
\hspace{0.3in}N_{\mu \nu }=\frac 12\left( \Psi _{\mu \nu }(x)-i \widetilde{
\Psi }_{\mu \nu }(x)\right).
\end{eqnarray*}

Adding and subtracting Eqs. (41), (42), we get Eqs. (38)-(40). The
self-dual tensor $M_{\mu \nu }$ which obeys Eqs. (41) is
transformed under $\left( 1,0\right) $ representation of the
Lorentz group and has three independent components. Equations (41)
are not invariant to the parity transformation and there is no
Lagrangian formulation of them. This also applies to Eqs. (42) for
the anti-self-dual tensor $N_{\mu \nu }$, which transforms under
$\left( 0,1\right) $ representation of the Lorentz group. But if
we consider the hole system of equations (41), (42) [which is
equivalent to Eqs. (38)-(40)] on the basis of $\left( 1,0\right)
\oplus \left( 0,1\right) $ representation of the Lorentz group, we
will have P-invariant theory within the Lagrangian formulation.

Consider now the first-order formulation of Eqs. (38)-(40) in
matrix form (10). Introducing a 14-dimensional wave function
\begin{equation}
\Psi (x)=\{\Psi _A(x)\}=\left(
\begin{tabular}{l}
$\psi _\mu (x)$ \\
$\Psi _{\mu \nu }(x)$ \\
$\widetilde{\psi }_\mu (x)$
\end{tabular}
\right), \label{43}
\end{equation}
where index $A=\mu $, $[\mu \nu ]$, $\widetilde{\mu }$ so that
$\Psi _\mu (x)\equiv \psi _\mu (x)$, $\Psi _{[\mu \nu ]}(x)\equiv
\Psi _{\mu \nu }(x)$, $\Psi _{\widetilde{\mu }}(x)\equiv
\widetilde{\psi }_\mu (x)$, we have that Eqs. (38)-(40) take the
form (10) with the matrices
\begin{equation}
\beta _\mu =\beta _\mu ^{(1)}+\beta _\mu ^{(2)}, \label{44}
\end{equation}
\begin{equation}
\beta _\mu ^{(1)}=\varepsilon ^{\lambda ,[\lambda \mu
]}+\varepsilon ^{[\lambda \mu ],\lambda }, \label{45}
\end{equation}
\begin{equation}
\beta _\mu ^{(2)}=\frac 12\epsilon _{\lambda \mu \alpha \beta }\left(
\varepsilon ^{\widetilde{\lambda },[\alpha \beta ]}+\varepsilon ^{[\alpha
\beta ],\widetilde{\lambda }}\right), \label{46}
\end{equation}
where $[\mu \nu ]$ means the antisymmetric combination of indexes
$\mu $ and $\nu $ and corresponds to the six-dimensional subspace
and there is a summation on the repeating indexes $\lambda $ in
(45), (46). Here the representation with the dimension
$2(6s+1)=14$ is valid. The $(6s+1)$ representation is built from
$M_\mu $ and self-dual tensor $M_{\mu \nu }$, or from $N_\mu $ and
the anti-self-dual tensor $N_{\mu \nu }.$ The resulting $2(6s+1)$
representation (43) is the direct sum of the above
representations. The fact that matrices (44) obey the algebra (12)
is confirmed by the equalities (3). This representation is
different from the $10(2s+1)$-dimensional representation (11). For
spin-1 particles, the wave function (11) has $10(2s+1)=30 $
components and is given by
\[
\Psi (x)=\left(
\begin{array}{c}
\Psi _{\alpha \beta }(x) \\
-\frac 1mD_\mu \Psi _{\alpha \beta }(x)
\end{array}
\right).
\]
The approach considered in Section 2 allows us to apply the unique
technique for arbitrary-spin particles which possess arbitrary
magnetic moment without introducing ``potentials'', but the cost
of this is the high dimension $ 10\left( 2s+1\right)$.

It should be noted that 10-dimensional matrices $\beta _\mu
^{(1)}$ in (45) obey the Petiau-Duffin-Kemmer algebra (7), and
that they act in 10-dimensional subspace of wave functions
\[
\Psi ^{PDK}(x)=\left(
\begin{array}{c}
\psi _\mu (x) \\
\Psi _{\mu \nu }(x)
\end{array}
\right),
\]
where we imply that $\Psi _{\mu \nu }(x)=D_\mu \psi _\nu (x)-D_\nu
\psi _\mu (x)$, so that 10-dimensional Petiau-Duffin-Kemmer
equation is given by
\begin{equation}
\left( \beta _\mu ^{(1)}D_\mu +m\right) \Psi ^{PDK}(x)=0.
\label{47}
\end{equation}

This corresponds to spin-1 particles without the doubling of spin states.

The generators of the Lorentz group in our 14-dimensional
representation (43) are given by
\begin{equation}
J_{\mu \nu }=\varepsilon ^{\mu ,\nu }-\varepsilon ^{\nu ,\mu
}+\varepsilon ^{[\lambda \mu ],[\lambda \nu ]}-\varepsilon
^{[\lambda \nu ],[\lambda \mu ]}+\varepsilon ^{\widetilde{\mu
},\widetilde{\nu }}-\varepsilon ^{\widetilde{ \nu },\widetilde{\mu
}}; \label{48}
\end{equation}
they obey the required commutation relations (17), (18). The
projection operators extracting states with definite 4-momentum
$p_\mu $ have virtually the same form (14), but with matrices
(44). The operator for the spin projection
\begin{equation}
\widehat{S}_{{\bf p}}=-\frac i{2\mid {\bf p\mid }}\epsilon
_{abc}p_aJ_{bc}, \label{49}
\end{equation}
with generators (48) satisfies the equation
\begin{equation}
\widehat{S}_{{\bf p}}\left( \widehat{S}_{{\bf p}}^2-1\right) =0.
\label{50}
\end{equation}

In accordance with this procedure [see (27)-(30)], we find the
projection operators corresponding to the definite spin
projections
\begin{equation}
\widehat{Q}_{\pm }=\frac 12\widehat{S}_{{\bf p}}\left(
\widehat{S}_{{\bf p} }\pm 1\right), \label{51}
\end{equation}
\begin{equation}
\widehat{Q}_0=1-\widehat{S}_{{\bf p}}^2. \label{52}
\end{equation}

Operators $\widehat{Q}_{\pm }$, $\widehat{Q}_0$ extract the spin
projections $s_k=\pm 1$ and $s_k=0$, respectively. Equations
(38)-(40) are invariant under the dual $SO(2)$ transformations
\[
\psi _\mu ^{\prime }(x)=\psi _\mu (x)\cos \alpha +\widetilde{\psi }_\mu
(x)\sin \alpha,
\]
\begin{equation}
\widetilde{\psi }_\mu ^{\prime }(x)=-\psi _\mu (x)\sin \alpha
+\widetilde{ \psi }_\mu (x)\cos \alpha. \label{53}
\end{equation}

This group of symmetry is connected with the doubling of spin
states of particles. To remove this degeneracy, we use the
projection generator (32) with the matrix
\begin{equation}
K=\varepsilon ^{\mu ,\widetilde{\mu }}+\varepsilon
^{\widetilde{\mu },\mu }+\frac 14\epsilon _{\mu \nu \alpha \beta
}\varepsilon ^{[\mu \nu ],[\alpha \beta ]}, \label{54}
\end{equation}

which has the property $K^2=I_{14}$, where $I_{14}$ is the unit matrix in the
14-dimensional space. Now we construct the density matrices of spin-1
particles for pure states as products of the projection operators:
\begin{equation}
\Pi _{\pm m,k,\varepsilon }=\frac 1{8m^2}\widehat{S}_{{\bf
p}}\left( \widehat{S}_{{\bf p}}+s_k\right) i\widehat{p}\left(
i\widehat{p}\pm m\right) \left( K+\varepsilon \right), \label{55}
\end{equation}
\begin{equation}
\Pi _{\pm m,0,\varepsilon }=\frac 1{4m^2}\left(
1-\widehat{S}_{{\bf p} }^2\right) i\widehat{p}\left(
i\widehat{p}\pm m\right) \left( K+\varepsilon \right), \label{56}
\end{equation}
where $\widehat{p}=p_\mu \beta _\mu$, and the spin projections on
the direction of the momentum ${\bf p}$ are{\bf \ }$s_k=\pm 1$ in
(55) and $s_k=0$ in (56), and $\varepsilon =\pm 1$. Matrices (55),
(56) can be represented in the form of matrix-dyads (35) with the
Hermitianizing matrix
\begin{equation}
\eta =\varepsilon ^{m,m}-\varepsilon ^{4,4}+\varepsilon
^{[m4],[m4]}-\frac 12\varepsilon ^{[mn],[mn]}-\varepsilon
^{\widetilde{m},\widetilde{m} }+\varepsilon
^{\widetilde{4},\widetilde{4}}, \label{57}
\end{equation}
where $m$, $n=1,2,3$, which obeys the general equations (20) with
matrices (44). The Lagrangian of free particles (where the vector
potential of the electromagnetic field $A_\mu =0$) takes the
standard form of
\[
{\cal L}=-\overline{\Psi }(x)\left( \beta _\mu \partial _\mu +m\right) \Psi
(x),
\]
where $\overline{\Psi }(x)=\Psi ^{+}(x)\eta $, and $\Psi ^{+}(x)$
is the Hermitian-conjugate wave function.

\section{Pair Production by External Electromagnetic Fields}

To calculate the probability of pair production of arbitrary-spin
particles, we follow the Nikishov method [22]. Thus, exact
solutions to Eq. (1) should be found for external, constant,
uniform electromagnetic fields. Using the properties of generators
$\Sigma _{\mu \nu }^{(\varepsilon )}$, we find the relationships

\begin{equation}
\frac 12\Sigma _{\mu \nu }^{(+)}F_{\mu \nu
}=S_iX_i,\hspace{1.0in}\frac 12\Sigma _{\mu \nu }^{(-)}F_{\mu \nu
}=S_iX_i^{*}, \label{58}
\end{equation}
where $X_i=H_i+iE_i$, $X_i^{*}=H_i-iE_i;$ $E_i,$ $H_i$ are the
electric and magnetic fields, respectively, and the spin matrices
$S_i$ obey Eqs. (2). In the diagonal representation, the equations
for the eigenvalues are

\begin{equation}
S_iX_i\Psi _{+}^{(\sigma )}(x)=\sigma X\Psi _{+}^{(\sigma )}(x),
\hspace{1.0in}S_iX_i^{*}\Psi _{-}^{(\sigma )}(x)=\sigma X^{*}\Psi
_{-}^{(\sigma )}(x), \label{59}
\end{equation}
where $X=\sqrt{{\bf X}^2},$ ${\bf X=H+}i{\bf E}$, and the spin
projection $ \sigma $ is

\begin{equation}
\sigma =
\begin{array}{c}
\pm s,\pm (s-1),\cdot \cdot \cdot 0 \\
\pm s,\pm (s-1),\cdot \cdot \cdot \pm \frac 12
\end{array}
\hspace{1.0in}
\begin{array}{c}
\mbox{for integer spins}, \\
\mbox{for half-integer spins}.
\end{array} \label{60}
\end{equation}

Taking into account (58) and (59), we rewrite Eq. (1) (for
$\varepsilon =\pm 1$) as

\begin{equation}
(D_\mu ^2-m^2-a\sigma X)\Psi _{+}^{(\sigma )}(x)=0,\hspace{0.5in}(D_\mu
^2-m^2-a\sigma X^{*})\Psi _{-}^{(\sigma )}(x)=0, \label{61}
\end{equation}
where $a=eq/s.$ These equations are like the Klein-Gordon equation
for scalar particles, but with complex ``effective'' masses
$m_{eff}^2=m^2+a \sigma X$ and $\left( m_{eff}^2\right)
^{*}=m^2+a\sigma X^{*}.$ It is sufficient to consider only one of
Eqs. (61). Let us consider the solution of the equation

\begin{equation}
(D_\mu ^2-m_{eff}^2)\Psi ^{(\sigma )}(x)=0,\hspace{1.0in}(\Psi ^{(\sigma
)}(x)\equiv \Psi _{+}^{(\sigma )}(x)) \label{62}
\end{equation}
in the presence of the external, constant, uniform electromagnetic
fields. The general case is when two Lorentz invariants of the
electromagnetic fields $ {\cal F}=\frac 14F_{\mu \nu }^2\neq 0$
and ${\cal G}=\frac 14F_{\mu \nu } \widetilde{F}_{\mu \nu }\neq 0$
($\widetilde{F}_{\mu \nu }=\frac 12\epsilon _{\mu \nu \alpha \beta
}F_{\alpha \beta }$; $\epsilon _{\mu \nu \alpha \beta }$ is the
antisymmetric Levi-Civita tensor). Then there is a coordinate
system in which the electric ${\bf E}$ and magnetic ${\bf H}$
fields are parallel, i.e., ${\bf E\parallel H}.$ In this case, the
4-vector potential is given by

\begin{equation}
A_\mu =\left( 0,x_1H,-x_0E,0\right), \label{63}
\end{equation}
so that 3-vectors ${\bf E}={\bf n}E$ and ${\bf H}={\bf n}H$ are
directed along the axes, where ${\bf n}=(0,0,1)$ is a unit vector.
The four solutions of Eq. (62) for the potential (63) with
different asymptotic forms are given in refs. [22, 37] (see also
refs. [38])

\begin{equation}
_{\pm }^{\pm }\Psi _{p,n}^{(\sigma )}(x)=N\exp \left\{ i(p_2x_2+p_3x_3)-
\frac{\eta ^2}2\right\} H_n(\eta )_{\pm }^{\pm }\psi ^{(\sigma )}(\tau ),\label{64}
\end{equation}
where $N$ is the normalization constant, $H_n(\eta )$ is the
Hermite polynomial,
\[
\eta =\frac{eHx_1+p_2}{\sqrt{eH}},\hspace{0.5in}\nu
=\frac{ik^2}{2eE}-\frac 12, \hspace{0.5in}\tau =\sqrt{eE}\left(
x_0+\frac{p_3}{eE}\right)
\]
and
\[
_{+}\psi ^{(\sigma )}(\tau )=D_\nu [-(1-i)\tau ],\hspace{1.0in}^{-}\psi
^{(\sigma )}(\tau )=D_\nu [(1-i)\tau ],
\]
\begin{equation}
^{+}\psi ^{(\sigma )}(\tau )=D_{\nu ^{*}}[(1+i)\tau ],\hspace{1.0in}_{-}\psi
^{(\sigma )}(\tau )=D_{\nu ^{*}}[-(1+i)\tau ]. \label{65}
\end{equation}

Here $D_\nu (x)$ is the Weber-Hermite function (the
parabolic-cylinder function). The probability for pair production
of particles with arbitrary spins by constant electromagnetic
fields can be obtained through the asymptotic form of the
solutions (65) when the time $x_0\rightarrow \pm \infty.$ At
$x_0\rightarrow \pm \infty $, the functions $_{+}^{+}\psi
^{(\sigma )}(\tau )$ have positive frequency and $_{-}^{-}\psi
^{(\sigma )}(\tau )$ have negative frequency. The constant $k^2$
which enters the index $\nu $ of the parabolic-cylinder functions
(65) is given by [38]

\begin{equation}
k^2=m_{eff}^2+eH(2n+1),\label{66}
\end{equation}
where $n=l+r,$ $l$ is the orbital quantum number, $r$ is the
radial quantum number, and $n=0,1,2,...$ is the principal quantum
number. It should be noted that for scalar particles, we have the
equation $k^2=p_0^2-p_3^2$, where $p_0$ is the energy and $p_3$ is
the third projection of the momentum of a scalar particle. In our
case of arbitrary-spin particles, the parameter $m_{eff}^2$ is a
complex value. Nevertheless all physical quantities in this case
are real values. Solutions (64), (65) are characterized by three
conserved numbers: $k^2$ and the momentum projections $p_2$,
$p_3.$ As shown in refs. [22], the functions (64) are connected by
the relations
\[
_{+}\Psi _{p,n}^{(\sigma )}(x)=c_{1n\sigma }{}^{+}\Psi _{p,n}^{(\sigma
)}(x)+c_{2n\sigma }{}^{-}\Psi _{p,n}^{(\sigma )}(x),
\]
\[
^{+}\Psi _{p,n}^{(\sigma )}(x)=c_{1n\sigma }^{*}{}_{+}\Psi _{p,n}^{(\sigma
)}(x)-c_{2n\sigma }{}_{-}\Psi _{p,n}^{(\sigma )}(x),
\]
\begin{equation}
^{-}\Psi _{p,n}^{(\sigma )}(x)=-c_{2n\sigma }^{*}{}_{+}\Psi _{p,n}^{(\sigma
)}(x)+c_{1n\sigma }{}_{-}\Psi _{p,n}^{(\sigma )}(x), \label{67}
\end{equation}
\[
_{-}\Psi _{p,n}^{(\sigma )}(x)=c_{2n\sigma }^{*}{}^{+}\Psi _{p,n}^{(\sigma
)}(x)+c_{1n\sigma }^{*}{}^{-}\Psi _{p,n}^{(\sigma )}(x),
\]
where coefficients $c_{1n\sigma },$ $c_{2n\sigma }$ are given by
\[
c_{2n\sigma }=\exp \left[ -\frac \pi 2(\lambda +i)\right] ,\hspace{1.0in}
\lambda =\frac{m_{eff}^2+eH(2n+1)}{eE},
\]

\[
\mid c_{1n\sigma }\mid ^2-\mid c_{2n\sigma }\mid
^2=1\hspace{1.0in} \mbox{for bosons},
\]

\begin{equation}
\mid c_{1n\sigma }\mid ^2+\mid c_{2n\sigma }\mid
^2=1\hspace{1.0in} \mbox{for fermions}.\label{68}
\end{equation}

The values $c_{1n\sigma },$ $c_{2n\sigma }$ are connected with the
probability of pair production of arbitrary-spin particles in the
state with the quantum number $n$ and the spin projection $\sigma
$. The absolute probability for the production of a pair in the
state with quantum number $n$, momentum $p$ and the spin
projection $\sigma $ throughout all space and during all time is

\begin{equation}
\mid c_{2n\sigma }\mid ^2=\exp \left\{ -\pi \left[ \frac{m^2}{eE}+\frac{
q\sigma H}{sE}+\frac HE(2n+1)\right] \right\}.\label{69}
\end{equation}

The value (69) is also the probability of the annihilation of a
pair with quantum numbers $n,$ $p,$ $\sigma $ with the energy
transfer to the external electromagnetic field. It is seen from
(69) that for $H\gg E$, the pair of particles are mainly created
by the external fields in the state with $n=0,$ $\sigma =-s.$ This
is the state with the smallest energy. So at $H\gg E$ there is a
production of polarized beams of particles and antiparticles with
the spin projection $\sigma =-s$ ( $s$ is the spin of particles).
The average number of pairs of particles produced from a vacuum is

\begin{equation}
\overline{N}=\int \sum_{n,\sigma }\mid c_{2n\sigma }\mid ^2dp_2dp_3\frac{L^2
}{(2\pi )^2} \label{70}
\end{equation}
because $(2\pi )^{-2}dp_2dp_3L^2$ is the density of final states,
where $ L$ is the cutoff along the coordinates, i.e., $L^3$ is the
normalization volume. The variables $\eta ,$ $\tau $ define the
region where the process occurs, which is described by solutions
(64) with the coordinates of the center of this region
$x_0=-p_3/eE,$ $x_1=-p_2/eH.$ Therefore instead of the integration
over $p_2$ and $p_3$ in (70), it is possible to use the
substitutions [22]

\begin{equation}
\int dp_2\rightarrow eHL,\hspace{1.0in}\int dp_3\rightarrow eET \label{71}
\end{equation}
with $T$ being the time of observation.

Evaluating the sum in (70) over the principal quantum number $n$, with the
help of (69), (71), we obtain the probability of pair production per unit
volume and per unit time

\begin{equation}
I(E,H)=\frac{\overline{N}}{VT}=\frac{e^2EH}{8\pi ^2}\frac{\exp
\left[ -\pi m^2/(eE)\right] }{\sinh \left( \pi H/E\right)
}\sum_\sigma \exp (-\sigma b), \label{72}
\end{equation}
where $b=\pi qH/(sE),$ $V=L^3.$ Now we evaluate the sum over the
spin projection $\sigma $ in (72) for integer and half-integer
spins:

1. Integer spins:
\[
\sum_\sigma \exp (-\sigma b)=S_1+S_{2,}\hspace{0.3in}
S_1=e^0+e^{-b}+...+e^{-sb}=\frac{e^{-b(s+1)}-1}{e^{-b}-1},
\]

\begin{equation}
S_2=e^b+e^{2b}+...+e^{sb}=\frac{e^b\left( e^{bs}-1\right)
}{e^b-1},S_1+S_2= \frac{\cosh (sb)-\cosh \left[ (s+1)b\right]
}{1-\cosh b}, \label{73}
\end{equation}

2. Half-integer spins:

\[
\sum_\sigma \exp (-\sigma b)=S_1^{^{\prime }}+S_2^{^{\prime }},\hspace{0.2in}
S_1^{^{\prime }}=e^{-b/2}+e^{-3b/2}+...+e^{-sb}=\frac{e^{-b(s+1)}-e^{-b/2}}{
e^{-b}-1},
\]

\begin{equation}
S_2^{^{\prime
}}=e^{b/2}+e^{3b/2}+...+e^{sb}=\frac{e^{b(s+1)}-e^{b/2}}{e^b-1}
,S_1^{^{\prime }}+S_2^{^{\prime }}=\frac{\cosh (sb)-\cosh \left[
(s+1)b\right] }{1-\cosh b}. \label{74}
\end{equation}

So the final expressions for integer and half-integer spins, (73)
and (74), are the same. Using the relationship

\begin{equation}
\frac{\cosh (sb)-\cosh \left[ (s+1)b\right] }{1-\cosh b}=\cosh
(sb)+\sinh (sb)\coth \frac b2=\frac{\sinh \left[ b(s+1/2)\right]
}{\sinh (b/2)} \label{75}
\end{equation}
and equations (72), (73) we arrive at the pair-production
probability

\begin{equation}
I(E,H)=\frac{\overline{N}}{VT}=\frac{e^2EH}{8\pi ^2}\frac{\exp
\left[ -\pi m^2/(eE)\right] }{\sinh \left( \pi H/E\right)
}\frac{\sinh \left[ (2s+1)q\pi H/(2sE)\right] }{\sinh \left[ q\pi
H/(2sE)\right] }. \label{76}
\end{equation}

Expression (76) coincides with that derived in refs. [23] using
the quasi-classical approach. So $I(E,H)$ is the intensity of the
creation of pairs of arbitrary-spin particles which possess the
magnetic moment $\mu =eq/(2m)$ and gyromagnetic ratio $g=q/s.$

In ref. [23], there is a discussion of physical consequences which
follow from Eq. (76). In particular, there is a pair production in
a purely magnetic field if $q=gs>1$ [23]. It is interesting that
the exact formula derived here from quantum field theory (which is
valid for arbitrary fields $E,$ $H$) coincides with the asymptotic
expression obtained in ref. [23] for $E,H\ll m^2/e.$

To obtain the imaginary part of the density of the Lagrangian, we
use the relationship [22]

\begin{equation}
VT\mbox{Im}{\cal L}=\frac 12\int \sum_{n,\sigma }\ln \mid
c_{1n\sigma }\mid ^2dp_2dp_3\frac{L^2}{(2\pi )^2}. \label{77}
\end{equation}

With the help of (68), (71), we arrive at (see also ref. [23])

\begin{equation}
\mbox{Im}{\cal L}=\frac{e^2EH}{16\pi ^2}\sum_{n=1}^\infty
\frac{\beta _n}n\exp \left( -\frac{\pi m^2n}{eE}\right)
\frac{\sinh \left[ n(2s+1)q\pi H/(2sE)\right] }{\sinh \left( n\pi
H/E\right) \sinh \left[ nq\pi H/(2sE)\right] }, \label{78}
\end{equation}
where
\[
\beta _n=
\begin{array}{c}
(-1)^{n-1}\hspace{1.0in}\mbox{for bosons}, \\
1\hspace{1.0in}\mbox{for fermions}.
\end{array}
\]

The different expressions for bosons and fermions occur due to
different statistics and relations (68). The first term ($n=1$) in
(78) coincides with the probability of the pair production per
unit volume per unit time divided by 2 [22] (see discussion in
ref. [23]).

\section{Vacuum Polarization of Arbitrary-Spin Particles}

Now we calculate the nonlinear corrections to the Lagrangian of a
constant, uniform electromagnetic field interacting with a vacuum
of arbitrary-spin particles with gyromagnetic ratio $g$. For the
case of spins $0$, $1/2$ and $1$ (for $g=2$) this problem was
solved in refs. [39,40,20,24]. The nonlinear corrections to the
Lagrangian of the electromagnetic field describe the effect of
scattering of light by light. We consider one-loop corrections to
the Maxwell Lagrangian corresponding to arbitrary-spin particles,
and, to take into account vacuum polarization, it is convenient to
adapt the Schwinger method [20]. Applying this approach to the
arbitrary spin particles described by Eq. (1), we arrive at the
effective Lagrangian of constant, uniform electromagnetic fields,

\begin{equation}
{\cal L}_1=\frac{\epsilon}{32\pi ^2}\int_0^\infty d\tau \tau
^{-3}\exp \left( -m^2\tau -l(\tau )\right) \mbox{tr}\exp \left(
\frac{eq}{2s}\Sigma _{\mu \nu }F_{\mu \nu }\tau \right),
\label{79}
\end{equation}
where $\epsilon =1$ for bosons, $\epsilon =-1$ for fermions,

\begin{equation}
\Sigma _{\mu \nu }=\Sigma _{\mu \nu }^{(+)}\oplus \Sigma _{\mu \nu
}^{(-)}, \hspace{0.5in}l(\tau )=\frac 12\mbox{tr}\ln \left[ \left(
eF\tau \right) ^{-1}\sin (eF\tau )\right] \label{80}
\end{equation}
and $F_{\mu \nu }$ is a constant tensor of the electromagnetic
field. The formal difference of (79) from the case of spin-$1/2$
particles is in the substitution $\sigma _{\mu \nu }\rightarrow
(q/s)\Sigma _{\mu \nu },$ where $ \sigma _{\mu \nu }=(i/2)\left[
\gamma _\mu ,\gamma _\nu \right] ,$ $\gamma _\mu $ being the Dirac
matrices. The problem is to calculate the trace of the matrices
occurring in the exponential factor in (79). Using relations
(58)-(60) and (73)-(75), we find

\begin{equation}
\mbox{tr}\exp \left( \frac{eq}{2s}\Sigma _{\mu \nu }F_{\mu \nu
}\tau \right) =2\mbox{Re}\left[ \cosh (eqX\tau )+\sinh (eqX\tau
)\coth \left( \frac{eqX\tau }{2s} \right) \right]. \label{81}
\end{equation}

Inserting (81) into (79) and adding the constant which is
necessary to cancel ${\cal L}_1$ when the electromagnetic fields
are turned off [20], we arrive at
\[
{\cal L}_1=\frac {\epsilon}{8\pi ^2}\int_0^\infty d\tau \tau
^{-3}\exp \left( -m^2\tau \right) \times
\]

\begin{equation}
\times \left[ (e\tau )^2{\cal G}\frac{\mbox{Re}\left[ \cosh
(eqX\tau )+\sinh (eqX\tau )\coth \left( eqX\tau /(2s)\right)
\right] }{2\mbox{Im}\cosh (eX\tau )}- \frac{2s+1}2\right],
\label{82}
\end{equation}
where ${\cal G}={\bf EH}$. With $q=1$ and $s=1/2$, this Lagrangian
(82) coincides with that of Schwinger [20]. Expression (82) is the
correction to the Maxwell Lagrangian that takes into account the
vacuum polarization of arbitrary- spin particles which possess the
magnetic moment $\mu =eq/(2m)$ and gyromagnetic ratio $g=q/s.$
Adding (82) to the Lagrangian of the free electromagnetic fields
\[
{\cal L}_0=-{\cal F=}\frac 12\left( {\bf E}^2-{\bf H}^2\right)
\]

and introducing the divergent constant for weak fields, we get the
expression for the total Maxwell Lagrangian
\[
{\cal L}_M={\cal L}_0+{\cal L}_1=-Z{\cal F}+\frac {\epsilon}{8\pi
^2}\int_0^\infty d\tau \tau ^{-3}\exp \left( -m^2\tau \right)
\times
\]

\begin{equation}
\times \left[ (e\tau )^2{\cal G}\frac{\mbox{Re}\left[ \cosh
(eqX\tau )+\sinh (eqX\tau )\coth \left( eqX\tau /(2s)\right)
\right] }{2\mbox{Im}\cosh (eX\tau )}- \frac{2s+1}2-4\beta (e\tau
)^2{\cal F}\right], \label{83}
\end{equation}
where

\begin{equation}
Z=1-\frac{\epsilon e^2\beta }{2\pi ^2}\int_0^\infty d\tau \tau
^{-1}\exp \left( -m^2\tau \right) ,\hspace{0.3in}\beta
=\frac{(2s+1)\left[ s\left( q^2-1\right) +q^2\right] }{24s}.
\label{84}
\end{equation}

Schwinger's procedure is used to renormalize the electromagnetic
field ${\cal F}\rightarrow Z{\cal F}$ and the charge $e\rightarrow
Z^{-1/2}e$. After expanding (84) in the small electric $E$ and
magnetic $H$ fields, we arrive at the Lagrangian of a constant,
uniform electromagnetic field (in rational units)

\begin{equation}
{\cal L}_M=\frac 12\left( {\bf E}^2-{\bf H}^2\right) -
\frac{2\epsilon\alpha ^2}{ 45m^4}\left[ \left( {\bf E}^2-{\bf
H}^2\right) ^2(15\beta -\gamma )+({\bf EH} )^2\left( 4\gamma
+\frac{2s+1}2\right) \right] +... \label{85}
\end{equation}
where $\alpha =e^2/(4\pi )$ and

\begin{equation}
\gamma =\frac{\left( 2s+1\right) \left[ q^4\left( s+1\right) \left(
3s^2+3s-1\right) -3s^3\right] }{16s^3}. \label{86}
\end{equation}

It is easy to verify that for the particular case of $s=1/2,$
$q=1$ (which corresponds to the Dirac theory), Eq.(85) coincides
with the well-known Schwinger Lagrangian [20]. With $s=1$ and
$q=2$ expression (85) is different from one obtained in ref. [24].
This is because our theory of particles with $s=1$ is not
equivalent to the Proca theory (see Section 3); there is the
doubling of spin states here. The effective Lagrangian (85) is of
the Heisenberg-Euler type, which has been found for the case of
the polarized vacuum of particles with arbitrary spins and
magnetic moment. Here we took into account virtual arbitrary-spin
particles, but not virtual photons. This is because at small
energies of the external fields, the radiative corrections are
small quantities. It is not difficult to find the asymptotic form
of (83) for supercritical fields $eE/m^2\rightarrow \infty $ and
$eH/m^2\rightarrow \infty$. It should be noted, however, that for
strong electromagnetic fields, the anomalous magnetic moment of
electrons depends on the external field [41,42] and hence there is
a similar dependence for arbitrary-spin particles. Therefore, to
obtain the correct limit, it is necessary to take into account
this dependence [26].

It is possible to obtain the imaginary part of the Lagrangian (78)
from (83) using the residue theorem, taking into account of the
poles of expression (83) and passing above them [20].

\section{Discussion of Results}

The theory of particles with arbitrary spins and magnetic moment
based on Eq. (1) and the corresponding Lagrangian allow us to find
density matrices (34),(35),(55), and (56), the pair-production
probability (76), and the effective Lagrangian for electromagnetic
fields (85), taking into account the polarization of the vacuum.
It is convenient to use matrix-dyads (34),(35),(55), and (56) for
different electrodynamic calculations in the presence of particles
with arbitrary spins. The exact formula for the intensity of pair
production of arbitrary-spin particles coincides with the
expression obtained in ref. [23] using the quasi-classical method
of ``imaginary time", which is valid only for $E,$ $H\ll m^2/e,$
i.e., for weak fields. Hence it follows that the analysis in ref.
[23] is valid for arbitrary electromagnetic fields and that it is
grounded in relativistic quantum field theory. In particular,
there is a pair production by a purely magnetic field if $gs>1$
[23], and in the presence of the magnetic field the probability
decreases for scalar particles and increases for higher spin
particles. As all divergences and the renormalizability are
contained in $\mbox{Re}{\cal L}$ (83), but not in $\mbox{Im}{\cal
L}$, the pair-production probability does not depend on the
renormalization scheme. The vacuum polarization corrections for
scalar and spinor (with $g=2$) particles are reliable because
their theories are renormalizable. The general formula (83) is of
interest for the further development of the field theory of
particles with higher spins (see discussion in refs. [23] and
[24]). Expression (83) is a reasonable result for arbitrary values
of $s$ and $q$ because for the particular case of scalar and
spinor particles, we arrive at known results. Thus, we have here a
reasonable and noncontradictory description of the nonlinear
effects that arise in this interaction.

\end{document}